\documentclass[showpacs,amsmath,twocolumn,showkeys,pra]{revtex4-1}
\usepackage{dcolumn}    % Align table columns on decimal point
\usepackage{bm}         % bold math
\usepackage{ifpdf}
\usepackage{amssymb,lineno,amsfonts}
\usepackage{graphicx}   % Include figure files
\usepackage{bbm}
\usepackage{mathrsfs}
\usepackage{upgreek}
\usepackage{epstopdf}
\usepackage{setspace}
\usepackage{hyperref}
\usepackage[matrix,frame,arrow]{xypic}
\usepackage{float}
\usepackage{natbib}
\usepackage{color} 
\definecolor{med-blue}{RGB}{25,25,112} 
\hypersetup{colorlinks, linkcolor={red},citecolor={blue}, urlcolor={blue}}

\newcommand{\ket}[1]{\vert{#1}\rangle}
\newcommand{\bra}[1]{\langle{#1}\vert}
\newcommand{\outpr}[2]{\vert{#1}\rangle\langle{#2}\vert}

\newcommand{\proj}[1]{\outpr{#1}{#1}}

\newcommand{\norm}[1]{\vert{#1}\vert^2}

\begin{document}
	
	\title{Spectral investigation of the noise influencing  multi-qubit states}
	\author{Deepak Khurana, Govind Unnikrishnan, and T. S. Mahesh}
	\email{mahesh.ts@iiserpune.ac.in}
	\affiliation{Department of Physics and NMR Research Center,\\
		Indian Institute of Science Education and Research, Pune 411008, India}
	
	\begin{abstract}
			{
Characterizing and understanding noise affecting quantum states  has immense benefits in spectroscopy as well as in realizing quantum devices.  Transverse relaxation times under a set of dynamical decoupling (DD) sequences with varying inter-pulse delays were earlier used for obtaining the noise spectral densities of single-qubit coherences.  In this work, using a pair of homonuclear spins and NMR techniques, we experimentally characterize noise in certain decoherence-free subspaces.  We also explore the noise of similar states in a heteronuclear spin-pair. Further using a 10-qubit system, we investigate noise profiles of various multi-qubit coherences and study the scaling of noise with respect to the coherence order.  Finally, using the experimentally obtained noise spectrum of the 10-qubit NOON state, we predict the performance of Uhrig DD sequence and verify it experimentally.
			}
	\end{abstract}
		
\maketitle

  %%%%%%%%%%%%%%%%%%%%%%%%% Introduction  %%%%%%%%%%%%%%%%%%%%5%%%%%%%%%%%%%%%%
 \section{Introduction}
  \label{Introduction}
  The inevitable presence of local or global electromagnetic noise may cause loss of quantum coherences of spin systems or induce redistribution of spin populations.  This phenomena which is often described in terms of decoherence or depolarization appears in NMR as a net relaxation of transverse or longitudinal magnetization.
  Combating decoherence is of utmost importance in spectroscopy and in realizing quantum devices such as  quantum information processors (QIP). Passive  techniques like decoherence-free subspaces (DFS) \cite{lidar1998decoherence,Lidar2003} as well as active techniques like dynamical decoupling (DD) \cite{viola1999dynamical} and quantum error correcting codes \cite{cory1998experimental,Preskill385} have been developed to overcome decoherence.  While the passive techniques rely on exploiting the symmetries in the interaction Hamiltonian, the active techniques focus on systematic modulation of the quantum states to suppress decoherence.  In the following we discuss the noise in various types of quantum coherences including DFS, single-quantum, as well as multiple-quantum coherences.
  
  An example of DFS is the singlet subspace in a two-qubit system \cite{Lidar2003}. In NMR, an excess population in the singlet state, over the uniformly distributed triplet states, is  termed as a singlet order.
  It has been shown that, such an order, under favorable circumstances, has much longer life-times than the usual longitudinal relaxation time scales, and is therefore known as a long-lived singlet state (LLS) \cite{carravetta2004long}.
  Similarly, the coherence between the singlet state and the zero-quantum triplet state also has longer life-times than the usual transverse relaxation time scales, and is therefore termed as a long-lived coherence (LLC) \cite{sarkarnmrs}. On the other hand, several other single- and multiple-quantum coherences lack the symmetry properties and are therefore prone to stronger decoherence \cite{jones2009magnetic}. 
  
  In this work we attempt to extract the noise spectra acting on various quantum coherences of NMR spin-systems.  Learning about noise-spectrum not only provides insights into the physical process of noise in quantum systems, but also assists in optimizing DFS conditions as well as in designing better controls for active suppression of noise.  
  Quantum noise spectroscopy (QNS), a tool to characterize the environmental noise, was independently proposed by Yuge et al \cite{yuge2011measurement} and \`Alvarez and Suter \cite{alvarez2011measuring}.
  
 The paper is organized as follows. In the following section we describe the theoretical formalism of QNS.  In section III, we apply QNS and experimentally extract the noise spectra of some interesting quantum coherences.  Finally we conclude in section IV.

  \section{Theory and Methods}
  \label{Theory}
  Here we review the theoretical aspects of characterizing the noise using a single two-level quantum system (qubit) as a probe.  We consider the qubit to be coupled to a bath via a purely dephasing interaction.
  Assuming the system Hamiltonian $\mathcal{H}_S = \omega_0 \sigma_z /2$ and the bath Hamiltonian $\mathcal{H}_B$, the joint-evolution is described by 
  the Hamiltonian
  \begin{eqnarray}
   \mathcal{H} = \mathcal{H}_S + \mathcal{H}_{SB} + \mathcal{H}_B.
   \end{eqnarray}
   Here $\mathcal{H}_{SB} = j_{SB} \sigma_z B/2 $ describes the system-bath interaction with $B$ being the bath  operator and $j_{SB}$ being the system-bath coupling strength.  In the interaction picture of the bath Hamiltonian, the bath operator 
   \begin{eqnarray}
   B'(t) = e^{-i\mathcal{H}_B t} B e^{i\mathcal{H}_B t}
   \end{eqnarray}
   becomes time-dependent.  
   After tracing-out the bath variables, the interaction Hamiltonian reduces to
   \begin{eqnarray}
   {\mathcal H}'_{S(B)} = j_{SB} b'(t)\sigma_z/2,
   \end{eqnarray}
    where $b'(t)$ is a stochastic function.
   We treat the bath to be classical and ${b}'(t)$ to be zero-mean stationary Gaussian process, as has been assumed before  \cite{yuge2011measurement,alvarez2011measuring}. However, an extension to a non-Gaussian case has also been reported recently \cite{PhysRevLett.116.150503}. 

  \begin{figure}
  	\hspace*{-0.4cm}
  	\includegraphics[trim = 0.6cm 5.8cm 0cm 5.6cm, clip, width=9.1cm ]{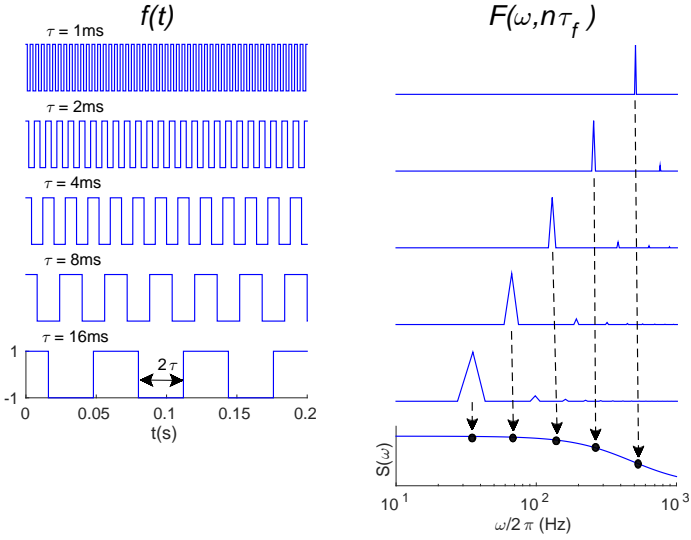}
  	\caption{The modulation functions $f(t)$ (left column) and the corresponding Fourier transforms, i.e., Filter functions $F(\omega,n\tau_f)$ (right column). 
 		The 1st order sampling points of the filter functions are illustrated using a schematic spectral density function $S(\omega)$ as shown in the lowest trace of the right column.}
 	\label{ff}
  \end{figure}   

    Suppose a DD sequence involving a series of $\pi$ pules is applied to refocus the dephasing caused by  $\mathcal{H}'_{S(B)}$. In the interaction representation associated with the DD sequence, the Hamiltonian $\mathcal{H}'_{S(B)}$  transforms to
    \begin{eqnarray}
\mathcal{H}_{S(B)}'' = f(t) j_{SB} b'(t)\sigma_z/2,
    \end{eqnarray} 
    where $f(t)$ is the modulation function that switches between $+1$ and $-1$ with the application of every $\pi$-pulse (see Figure \ref{ff}). The Fourier transform of $f(t)$ is known as the Filter function $F(\omega,\tau_{f})$, where $\tau_{f}$ is the period of $f(t)$.
    
    Noise spectral density $S(\omega)$ is defined as the Fourier transform of the autocorrelation function $g(\tau) = \langle b'(\tau)b'(t+\tau)\rangle$.  
      The decay of quantum coherence is influenced by the noise spectral density as well as the filter function \cite{Abragam,Biercuk}.  This decay can be modeled in the form $\exp(-t R(t))$, wherein the time-dependent argument for $n$ iterations of a DD sequence is given by,
    \begin{equation}
    \label{PSDint}
    t R(t) =  \sqrt{\frac{\pi}{2}} \int_{-\infty}^{\infty} d\omega S(\omega) |F(\omega,n\tau_f)|^2,
    \end{equation}
    \cite{PhysRevLett.87.270405,PhysRevLett.93.130406,PhysRevB.77.174509}.

	Fourier analysis of DD induced modulations is similar to that of a diffraction grating.
     For time-scales much larger than the noise correlation times, i.e., for large $n$, the filter-function  becomes time independent and reduces to a delta-comb
     \begin{align}
    \norm{F(\omega,n\tau_f)} = \sum_{k=-\infty}^\infty \frac{n\sqrt{2\pi}}{\tau_f} \delta(\omega-\omega_k)
    \norm{F(\omega,\tau_f)}
     \end{align} 
     where  $\omega_k = 2\pi k/\tau_{f}$ and $k\in[-\infty,\infty]$ is the Fourier index of $f(t)$ \cite{PhysRevA.83.032303}.  The exponential decay factor now becomes time independent, i.e., $R = 1/T_2$. Hence for a long time point $t = n\tau_f$ 
     \begin{eqnarray}
     \frac{1}{T_2} &=& \frac{2\pi}{\tau_f^2}\sum_{k=0}^{\infty}
      S(\omega_k) \norm{F(\omega_k,\tau_f)} \nonumber \\
     &=&\sum_{k=0}^{\infty}S(\omega_k) A_k^2,
     \label{eq7}
     \end{eqnarray}  
      where\ $A_k^2 =  \frac{{2\pi}}{\tau_{f}^2}|F(\omega_k,\tau_{f})|^2$
     \cite{alvarez2011measuring}.
     
    In the case of a free-evolution without any DD sequence, the modulation function $f(t)$ becomes constant and therefore, the filter-function $F(\omega,\tau_f)$ is a sinc-function centered at $\omega=0$, and the decay rate $1/T_2$ depends only on $S(0)$.
    
For the CPMG sequence  \cite{PhysRev.94.630,meiboom1958modified}
    with uniformly distributed $\pi$ pulses at an interval $2\tau$,     $f(t)$ switches between $+1$ and $-1$ with a period $\tau_{f} = 4\tau$.
    The schematic diagrams of $f(t)$ and the corresponding filter functions $\norm{F(\omega,\tau_f)}$ for a set of $\tau$ values are shown in Fig. \ref{ff}. 
    In this case, $A_k^2 = (4/{\pi}^2k^2)$ for odd $k$ and $A_k = 0$ otherwise. Hence
    \begin{equation}
    \label{PSD}
    \frac{1}{T_2} = \frac{4}{\pi^2}\sum_{l=0}^{\infty} \frac{1}{(2l+1)^2}\ S(\omega_{2l+1}).
    \end{equation}
Thus the decay rate $1/T_2$ for a given $\tau$ is determined by the harmonics at $\omega_{2l+1} = \pi(2l+1)/2\tau$, as illustrated in Fig. \ref{ff}.
Hence from the experimentally measured $T_2$ values for  $\tau \in [\tau_\mathrm{min},\tau_\mathrm{max}]$, one can extract the spectral density points $S(\omega_{2l+1})$ in the range 
$\omega \in[ \pi/2\tau_\mathrm{max}, \pi/2\tau_\mathrm{min}]$
 by inverting the above equation.  In the following we discuss two ways of extracting the noise spectrum $S(\omega)$ from Eq. \ref{PSD}.

An approximate way is to truncate the series in Eq.  \ref{PSD} to the zeroth order term so that,
% One way to achieve this is by approximating the above expression to keep only the first order term if the spectral density decreases sharply with $\omega$ \cite{yuge2011measurement}
\begin{equation}
S\left(\frac{\pi}{2\tau}\right) \approx  \frac{\pi^2}{4T_2}.
\end{equation}
This method is suitable for spectral densities with sharp cut-offs at low-frequencies \cite{Swathi}.  Otherwise, ignoring higher order terms may introduce an error up to about 10\%. 

On the other hand, we can account for the zeroth as well as many higher order terms of spectral density by using a suitable model function for the spectral density.
Random isotropic rotations of liquid molecules usually lead to exponential autocorrelation function  and therefore the corresponding spectral density is Lorentzian \cite{Abragam}.  Multiple relaxation sources may lead to multi-Lorentzian spectral density as observed in the experiments described in the next section.  
Our phenomenological model thus consists of a linear combination of Lorentzians
\begin{equation}
S_L(\omega) = \sum_{j=1}^{L} \frac{\lambda_j}{(\omega-\omega_j)^2+\lambda_j^2}.
\label{multilor}
\end{equation}
The parameters $\omega_j$ (center-frequency) and $\lambda_j$ (line-width) can be determined by numerically maximizing the overlap between the experimental $T_2$ values and those calculated using the model function $S_L(\omega)$. 
Another benefit of obtaining the functional form of spectral density is that it allows one to evaluate the performance of various DD sequences at arbitrary inter-pulse spacing, as illustrated in section III C. 

Although, noise filtering techniques for multi-qubit states are being developed recently \cite{su2012filter,viola2016}, in this work we use a single probe qubit to capture effective noise influencing multiqubit states.
 
  \section{Experiments and results}
  \label{Methods}
  In this section, we describe the experimental noise spectroscopy of certain interesting multi-qubit  coherences.
  
  \subsection{LLS and LLC}
  \label{LLS and LLC}
  We used the two phenyl $^{1}$H nuclei of  2,3,6-trichlorophenol dissolved in dimethyl sulphoxide-D6.  The experiments were carried out at 300 K in two different magnetic fields corresponding to Larmor frequencies $\nu_0=400$ MHz as well as $\nu_0 = 600$ MHz.
    The chemical shift difference $\Delta \nu \times 10^6/\nu_0 = 0.21$ ppm and the scalar coupling constant $J = 8$ Hz. 
    Under weak-coupling approximation, the NMR Hamiltonian is
    \begin{equation}
    \label{ham1}
    \mathcal{H}= \pi\Delta\nu I_z-\pi\Delta\nu S_z+ \pi J 2 I_z S_z,
    \end{equation}
    where $I_z$ and $S_z$ are the spin operators.

  \begin{figure}[b]
  	\includegraphics[trim = 0.5cm 4cm 6.9cm 0.5cm, clip, width=8.7cm ]{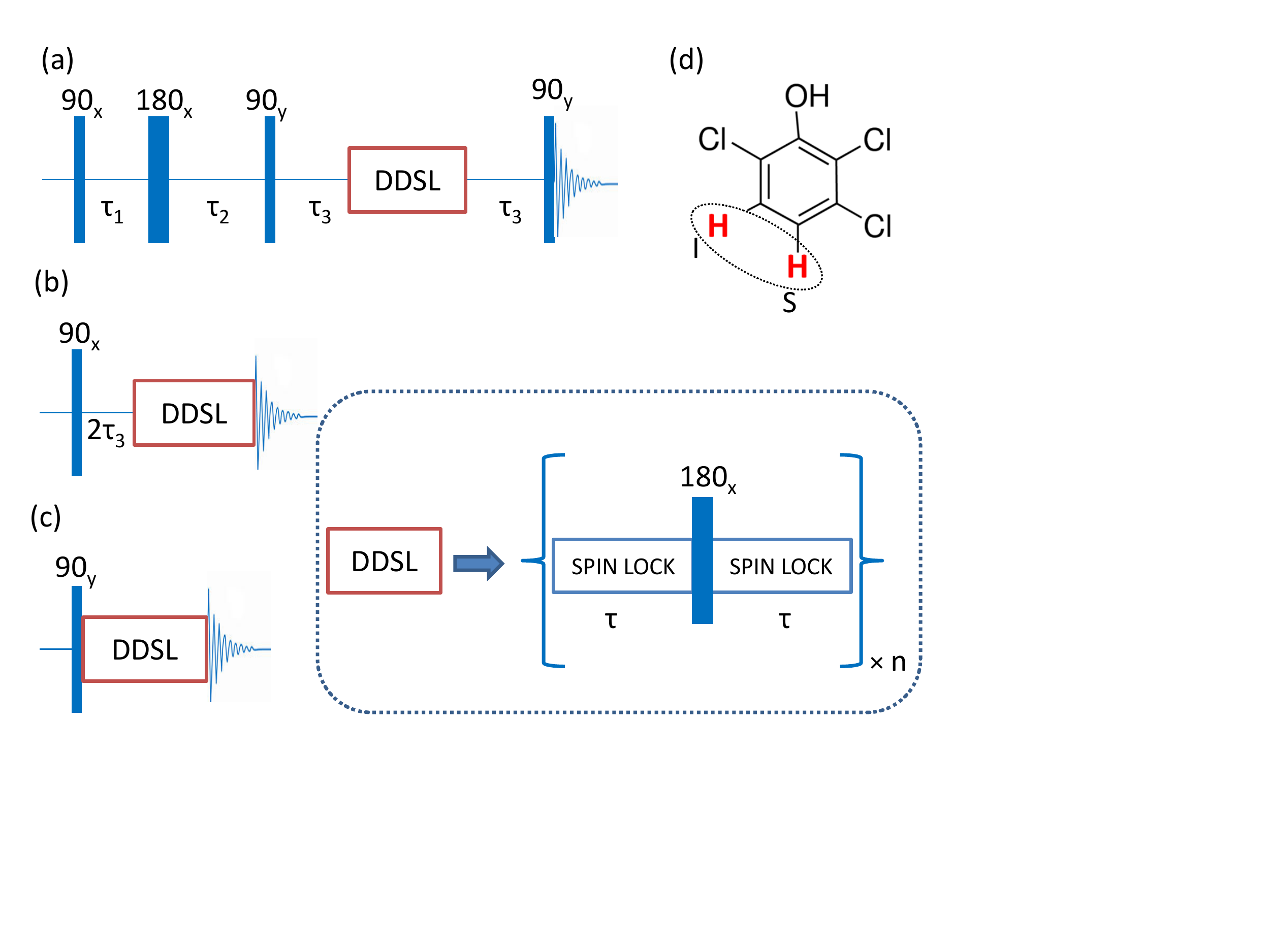}
  	\caption{Pulse sequences used to measure the noise spectrum of (a) $\rho_\mathrm{{_{LLS}}}$, (b) $\rho_\mathrm{{_{LLC}}}$, and (c) $\rho_\mathrm{{_{SL}}}$. Here $\tau_1 = 1/(4J)$, $\tau_2 = 1/(4J)+1/(2\Delta \nu)$, $\tau_3 = 1/(4 \Delta \nu) $, and $n$ is the number of times the loop is repeated. (d) Structure of 2,3,6-trichlorophenol.  The CPMG DD sequence with spin-lock along x-axis is shown in the inset (DDSL). } 	\label{DFSpulseq}
  \end{figure}
     
   The natural choice for expressing LLS and LLC is the singlet triplet basis, formed by the eigenvectors of the isotropic interaction Hamiltonian $\mathbf{I} \cdot \mathbf{S}$, i.e.,
  \begin{align}
  \ket{T_0}&=\frac{1}{\sqrt{2}}(\ket{01}+\ket{10}),  \nonumber \\
  \ket{T_{+}}&=\ket{00}, \nonumber \\
  \ket{T_{-}}&=\ket{11}, ~\mbox{and} \nonumber \\
  \ket{S_0}&=\frac{1}{\sqrt{2}}(\ket{01}-\ket{10}) \label{singlet},
  \end{align}
   where $\{\ket{00}, \ket{01}, \ket{10}, \ket{11} \}$ form the Zeeman eigenbasis.

In particular, we focus on the following coherences:
    \begin{align}
        \label{LLS}
        \rho_\mathrm{{_{LLS}}}&=\ket{S_0} \bra{S_0}-\ket{T_0}\bra{T_0} \nonumber \\
    \rho_\mathrm{{_{LLC}}}&=\ket{S_0} \bra{T_0}+\ket{T_0}\bra{S_0} \nonumber \\
    \rho_\mathrm{_{SL}}&=\proj{T_+}-\proj{T_-}.
        \end{align}
In the above, ${\rho}_\mathrm{_{LLS}}$,
$\rho_\mathrm{_{LLC}}$, and $\rho_\mathrm{_{SL}}$ are realized by preparing
the states $-\mathbf{I} \cdot \mathbf{S}$, $I_x-S_x$, and $I_x+S_x$
respectively, and applying a strong spin-lock along the $x$ axis \cite{carravetta2004long,sarkarnmrs}.
Here we have considered  $\rho_\mathrm{_{SL}}$ for the sake of comparison with the other long-lived states.  The pulse sequences corresponding to these states are shown in Fig. \ref{DFSpulseq}.

\begin{figure}[h]
  	\includegraphics[trim = 1cm 6.8cm 1.5cm 7cm, clip, width=9cm ]{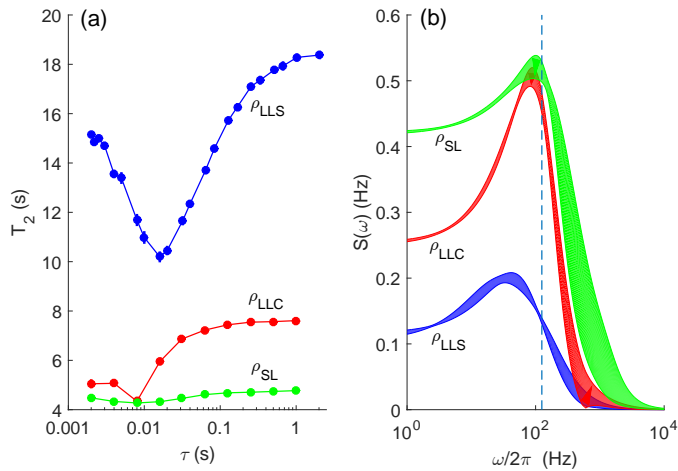}
  	\caption{(a) Experimental decay constants (dots) of 2,3,6-trichlorophenol (averaged for both protons) in 400 MHz spectrometer for a range of $\tau$ values and for different states as indicated.  The solid lines correspond to decay constants obtained from the best fit by the 3-Lorentzian model as described in Eq. \ref{multilor}. (b) The corresponding noise spectral density bands.  The dashed line at 125 Hz corresponds to the maximum harmonics sampled with $\tau=2$ ms. }
  	\label{dfs400}
  \end{figure}

  We use the multi-Lorentzian model function described in Sec. \ref{Theory} to extract the noise spectrum. 
   The best fit was achieved with a minimum of three Lorentzian functions (i.e., $L=3$) as described in Eq. \ref{multilor}.  
   We scan over a range of spectral frequencies $\omega=\pi/2\tau$ by varying the duration $2\tau$ between the $\pi$ pulses, and measure the corresponding $T_2$ values.  WALTZ-16 spin-lock of 2 kHz amplitude was applied along the x-axis during the delays between the $\pi$ pulses. 
%  To avoid phase artifacts due to J-evolution, we choose the iteration number $n$ such that the total duration of DD is an integral multiple of $1/J$, i.e., $ 2n\tau J $ is an integer.  
  The experimental $T_2$ values for all the three states and for $\tau$ values ranging from 2 ms to 2 s are displayed in Fig. \ref{dfs400}(a).  
  The uncertainties in the noise spectrum (represented by the width of the bands) are estimated by several iterations of maximizations also considering the standard deviations in $T_2$ values.
  
   As expected, $\rho_\mathrm{_{LLS}}$ has the lowest noise in the whole-frequency range indicating long-lifetimes.  On the other hand, $\rho_\mathrm{_{SL}}$ has the highest noise indicating a relatively short-lived state.  The long-lived coherence $\rho_\mathrm{_{LLC}}$ has an intermediate noise-profile.  Owing to the hardware limitations, the highest frequency sampled by the experiments is 125 Hz (indicated by a dotted line in Fig. \ref{dfs400}(b)), corresponding $\tau = 2$ ms.  The noise-profiles above this cutoff frequency are basically an extrapolation obtained by the model functions.
    Interestingly, in all the three spectral-density profiles we observe a hump close to 100 Hz.  Replacing the hydroxyl proton with deuterium did not affect the hump.  We have also observed a systematic dependence of the hump with the spin-lock power, which possibly relates its origin to an interference between spin-lock and DD sequences.  However further investigations are required to confirm this point.

      \begin{figure}[t]
      	\includegraphics[trim = 0.6cm 7.6cm 0.8cm 8cm, clip, width=9cm ]{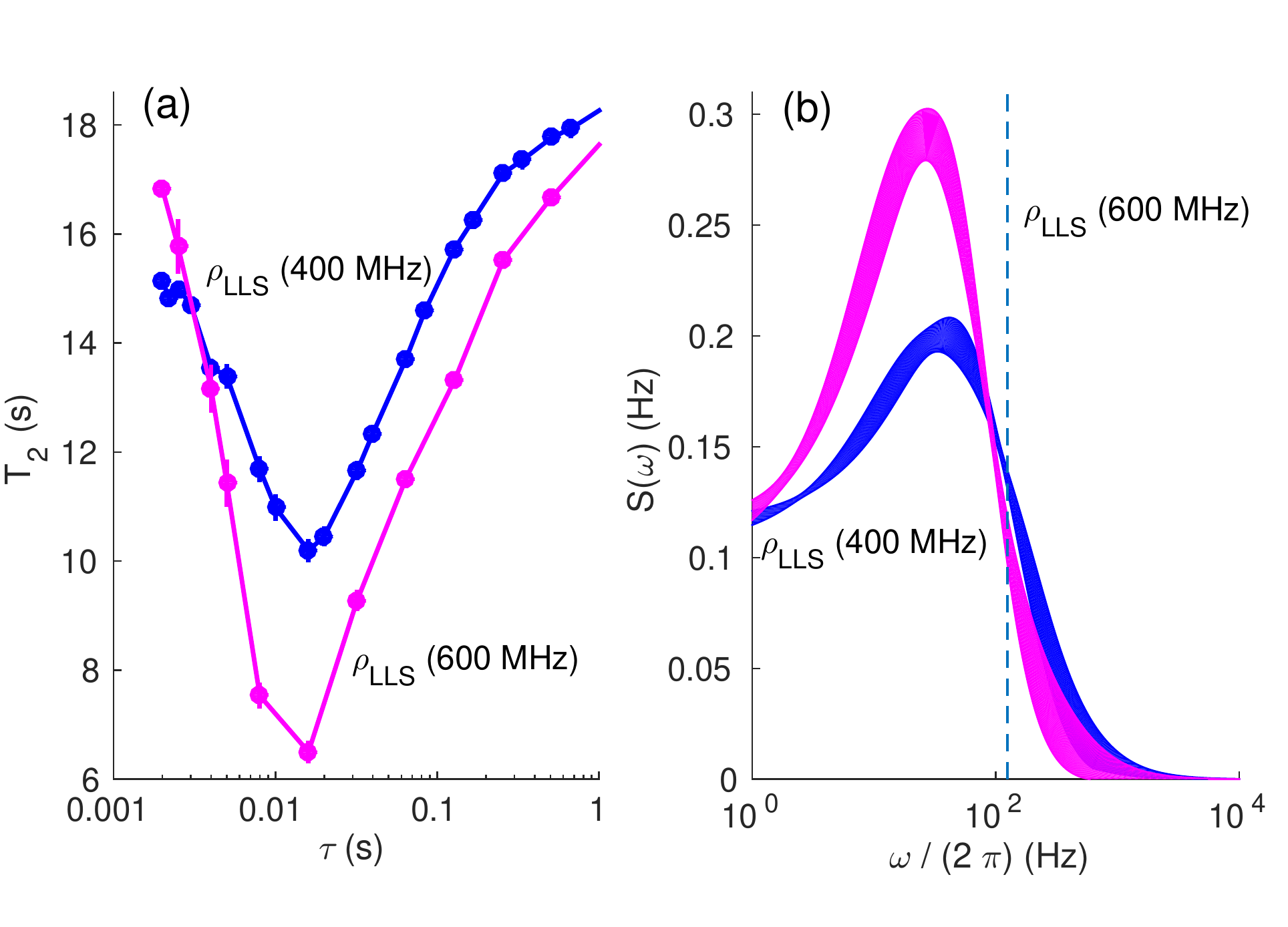}
      	\caption{(a) The experimental decay constants (dots) at various $\tau$ delays for the singlet state $\rho_\mathrm{_{LLS}}$ at two different magnetic fields, i.e., 400 MHz and 600 MHz as indicated. The solid lines correspond to decay constants obtained from the best fit by the 3-Lorentzian model as described in Eq. \ref{multilor}.  (b) The corresponding noise spectra.}
      	\label{600vs400}
      \end{figure}

Although it is well known that singlet state is longer lived at lower fields\cite{pileio2009theory1,pileio2009theory2}, it is not obvious how the spectral characteristics  of noise changes under a higher field.   Therefore it is useful to compare the noise spectrum at two different fields.  With this intention, we have measured the noises of singlet state of same system, i.e., the proton pair of 2,3,6-trichlorophenol, at 400 MHz as well as at 600 MHz spectrometers under identical conditions.  The $T_2$ values and the corresponding spectral density bands are shown in Fig. \ref{600vs400}.  As expected, the noise is significantly stronger at 600 MHz.

  \begin{figure}[b]
  	\includegraphics[trim =3.3cm 4.5cm 3cm 4.2cm, clip, width=8.5cm ]{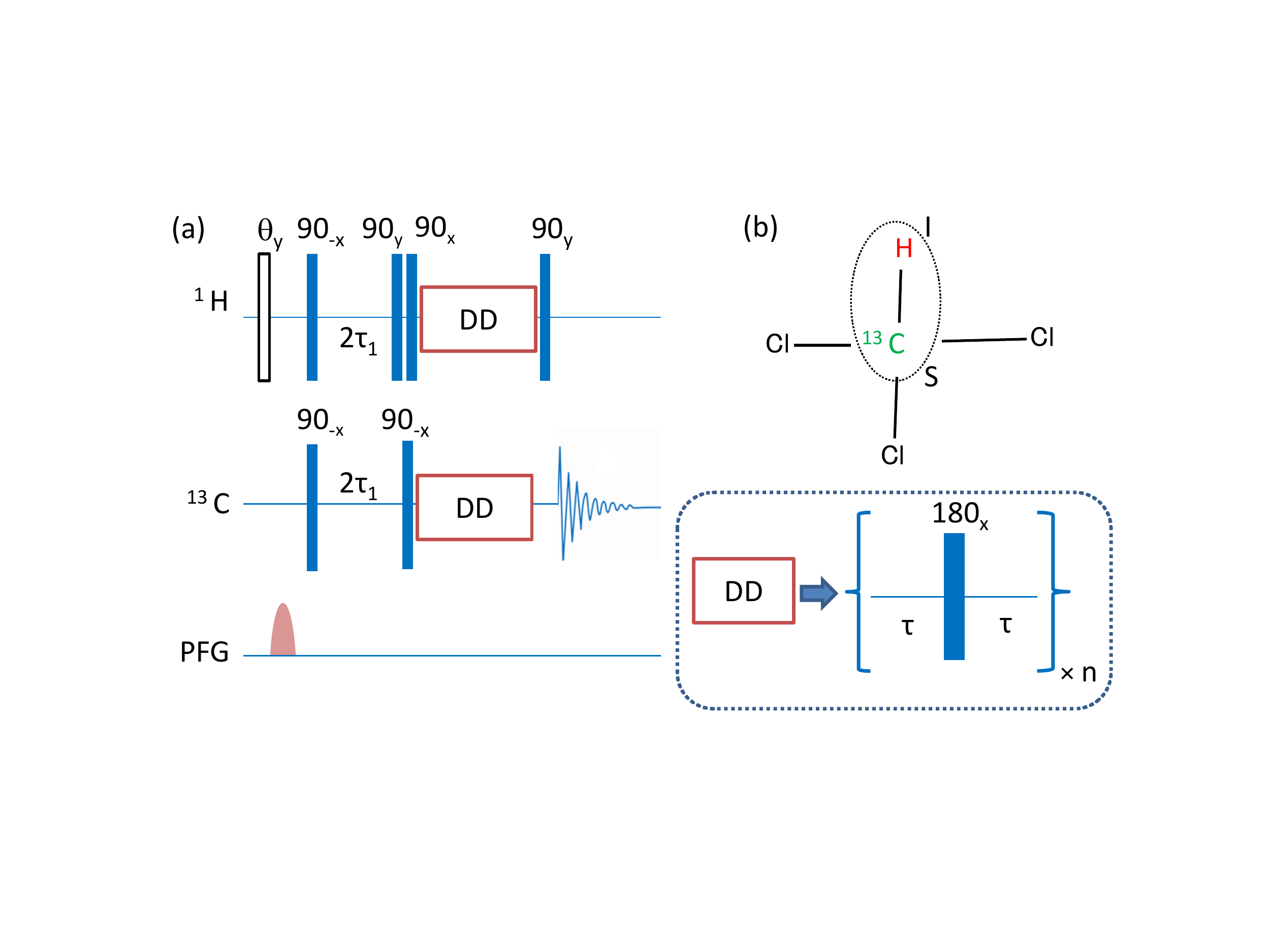}
  	\caption{(a) Pulse sequence to measure noise spectrum in a heteronuclear spin system. (b) Molecular structure of Chloroform. Here singlet state is prepared on  $^{1}$H and $^{13}$C spins with a coupling constant $J_\mathrm{CH} = 209$ Hz between them. The CPMG DD sequence  is shown in the inset.}
  	\label{chcl3}
  \end{figure}
  
   \begin{figure}[t]
   	\includegraphics[trim = 0.5cm 7cm 1.5cm 6.7cm, clip, width=8.8cm ]{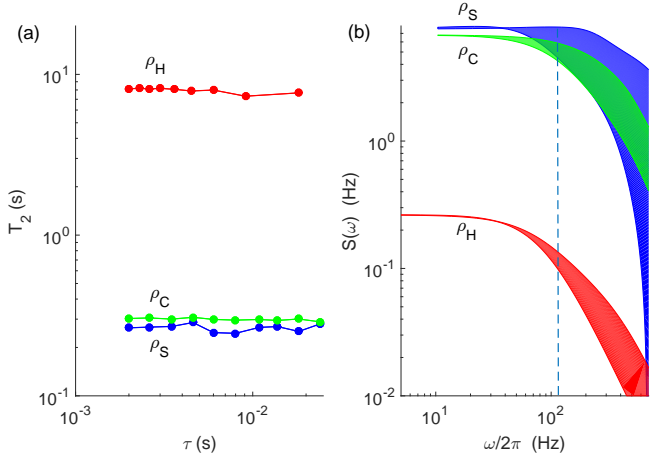} 
   	\caption{(a) The experimental decay constants (dots) of $^{13}$C-Chloroform at various $\tau$ delays for single spin states $\rho_H=I_x$, $\rho_C = S_x$, and the singlet state $\rho_S$ at 500 MHz. 
   		The solid lines correspond to decay constants obtained from the best fit by the 3-Lorentzian  model as described in Eq. \ref{multilor}.
   		 (b) The corresponding noise spectra.}
   	\label{nschcl3}
   \end{figure}
   
  \subsection{Heteronuclear spin-pair}
  In a hetronuclear spin pair, such as $^1$H-$^{13}$C in $^{13}$C Chlorform (dissolved in CDCl$_3$; see Fig. \ref{chcl3}b), the singlet subspace is not a DFS, because a strong magnetic field breaks the symmetry between two spins and a spin-lock to restore the symmetry is not practical.  Therefore,  a heteronuclear singlet-state, though easy to prepare, is no longer an eigenstate of the interaction Hamiltonian. A pulse sequence to measure their noise spectrum is shown in Fig. \ref{chcl3}a. It begins with a $\theta = \cos^{-1}(1/4)$ pulse on $^1$H spin followed by a pulsed-field-gradient to equalize the polarizations and prepare the state $I_z+S_z$.  The following RF pulses and delays convert it to $-I_xS_x-I_yS_y \equiv \proj{S_0}-\proj{T_0}$.  A CPMG DD sequence with a variable $\tau$ delay followed by a final $90_y$ on $^1$H is then used to measure the noise spectrum.  The results are shown in Fig. \ref{nschcl3}.
  For comparison, we have also included the noise spectra of single-spin states $I_x$ and $S_x$.  Here  $^{1}$H  spin has longer $T_2$ values and accordingly lower noise profile compared to $^{13}$C.  Unlike in the homonuclear case, the heteronuclear singlet has the shortest $T_2$ values and therefore highest noise profile.  Therefore a heteronuclear singlet is not an LLS at high fields \cite{levitt2012singlet}.

  \subsection*{C. Large quantum coherences}
  \label{MQE}
\begin{figure}
 	\includegraphics[trim = 4.6cm 5.7cm 7.3cm 2.6cm, clip, width=7.5cm]{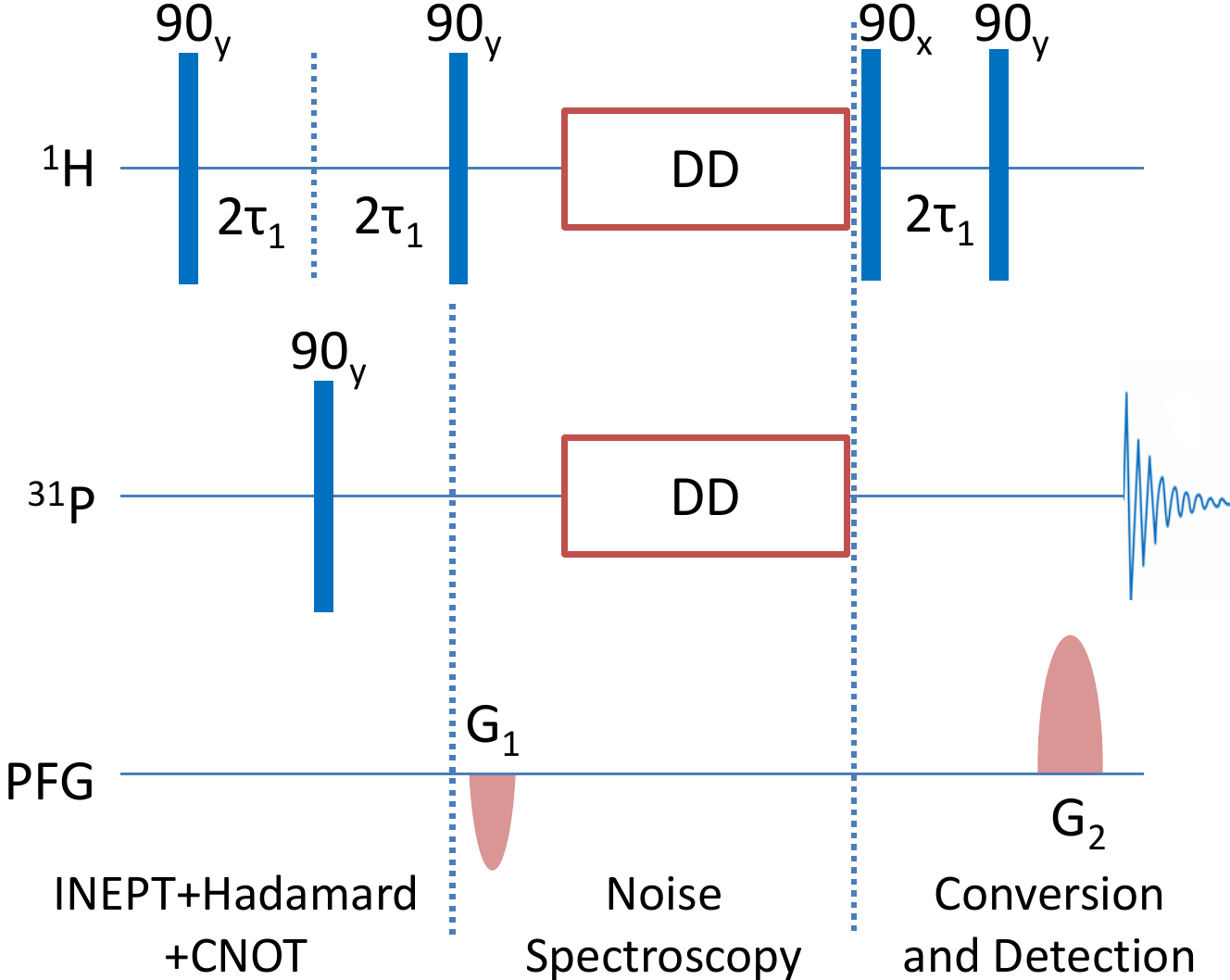}
 	\caption{Pulse sequence to measure noise spectra of the MSSM states. An initial INEPT (insensitive nuclei enhanced by polarization transfer) \cite{Cavanagh} operation is used to transfer magnetization from $^{1}$H to $^{31}$P. The PFGs $G_1$ and $G_2$ are chosen such that $\phi_2(k) = -\phi_1(k)$ to select out an MSSM state with a particular lopsidedness $l(k)$. A CPMG-DD sequence with composite $\pi$-pulses was used. }
 	\label{MQCp}
 \end{figure}

  Consider an $N$-spin star-topology system wherein a central spin (denoted by $M$) is uniformly coupled to $N-1$ magnetically equivalent spins (denoted by $A$).  Such a system allows a convenient way to prepare many large quantum coherences.  The method involves applying a Hadamard gate (denoted by H) on the central spin followed by a CNOT gate as described in Fig. \ref{MQCp}.  In thermal equilibrium, the central spin will have an excess $\ket{0}_M$ population while the surrounding spins have a Boltzmann distribution over all the states $\ket{N-1,0}_A$ to $\ket{0,N-1}_A$, wherein the first and second numbers denote the numbers of spins in $\ket{0}$  and $\ket{1}$ states respectively.  The effect of Hadamard and CNOT gates can now be described as
  \begin{eqnarray}
 \ket{0}_M\sum_{k=0}^{N-1} \ket{N-1-k,k}_A 
  \stackrel{H}{\rightarrow} & \nonumber \\
  \frac{\ket{0}_M+\ket{1}_M}{\sqrt{2}}\sum_{k=0}^{N-1} \ket{N-1-k,k}_A  \stackrel{CNOT}{\longrightarrow} & \nonumber \\
  \frac{1}{\sqrt{2}}\sum_{k=0}^{N-1}  \ket{0}_M\ket{N-1-k,k}_A + \ket{1}_M\ket{k,N-1-k}_A. & \nonumber   
  \end{eqnarray}
  The last sum represents a collection of coherences with quantum numbers $N,N-2,\cdots, 0$ for even $N$ and $N,N-2,\cdots, 1$ for odd $N$.  Such coherences are often referred to as $\ket{\mathrm{MSSM}}$ (many-some, some-many) states \cite{jones2009magnetic}.  
  A special MSSM state is the $N$-quantum $\ket{\mathrm{NOON}}$ state 
    \begin{equation}
    \ket{\mathrm{NOON}} = (\ket{000..0} +   \ket{111..1})/\sqrt{2}.
    \end{equation}

  The MSSM states can be individually studied by selective filtering of their signals using a pair of pulsed-field-gradients (PFG) (see Fig. \ref{MQCp}).  If $\gamma_A$ and $\gamma_M$ denote the respective gyromagnetic ratios of $A$ and $M$ spins, we can express the dephasing caused by the first PFG by,
  \begin{eqnarray}
  \phi_1(k) \propto \frac{\gamma_M + (N-2k-1)\gamma_A}{\gamma_A} = l(k),
  \end{eqnarray}
  where the term in the right hand side is known as the lopsidedness of the MSSM state and 
  \begin{eqnarray}
   \frac{\gamma_M-(N-1)\gamma_A}{\gamma_A} \le l(k) \le \frac{\gamma_M+(N-1)\gamma_A}{\gamma_A}. 
  \end{eqnarray}
  
     \begin{figure}[t]
     	\includegraphics[trim = 4cm 2.6cm 3.5cm 1.5cm, clip, width=7.5cm]{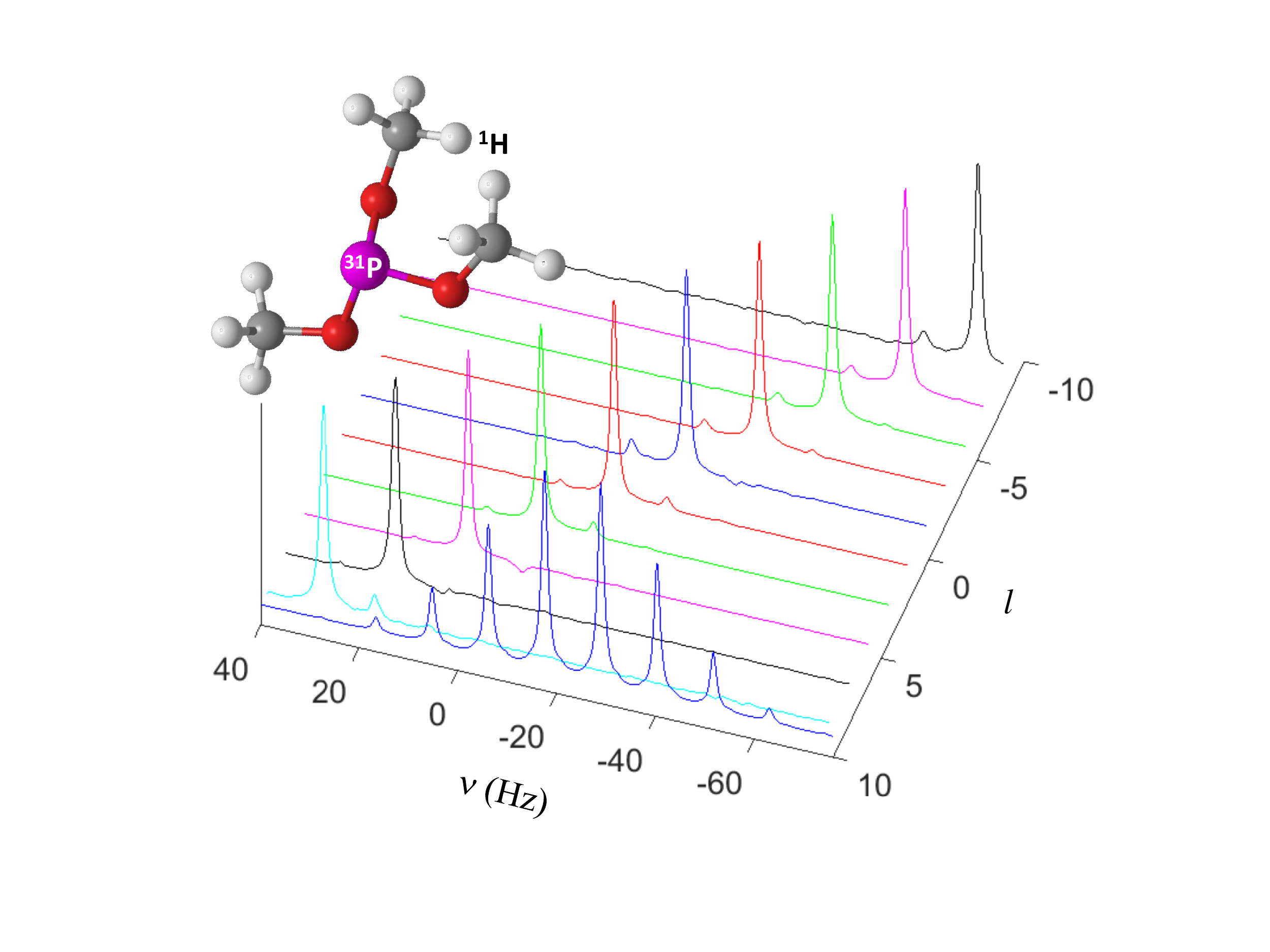}
     	\caption{The spectral lines corresponding to various MSSM states with varying lopsidedness $l$.  Each spectral line is individually normalized.  The reference spectrum with all the lines is shown at the front.  The structure of Trimethylphosphite is also shown at the top-left corner. }
     	\label{TMP}
     \end{figure}
  
  Each MSSM state is converted back into an observable single-quantum $M$ spin coherence by the application of a second CNOT
  \begin{eqnarray}
%  \hspace*{-1cm}
 \frac{1}{\sqrt{2}}\sum_{k=0}^{N-1}  \ket{0}_M\ket{N-1-k,k}_A + e^{i\phi_1(k)} \ket{1}_M\ket{k,N-1-k}_A
 \nonumber \\
%   \hspace*{-1cm}
  \stackrel{CNOT}{\longrightarrow} 
   \left(\sum_{k=0}^{N-1}\frac{\ket{0}_M+e^{i\phi_1(k)}\ket{1}_M}{\sqrt{2}}\right)  \ket{N-1-k,k}_A. \nonumber
   \end{eqnarray}

Selection of the signal from a desired MSSM state with a particular $l(k)$ value is achieved with the help of a second PFG which introduces a phase $\phi_2(k) = -\phi_1(k)$.  The noise spectroscopy of the MSSM states can be studied by inserting the DD sequence just before the second CNOT (see Fig. \ref{MQCp}).

  Experiments were carried out in a Bruker 500 MHz spectrometer at 300 K. Trimethylphosphite (see Fig. \ref{TMP}) dissolved in DMSO was used as a 10-spin star-topology system including a central $^{31}$P spin ($M$ spin) and the nine surrounding $^{1}$H spins ($A$ spins).  The scalar spin-spin coupling $J_\mathrm{PH}$ was about 11 Hz.  The signals from various MSSM states (obtained with the pulse-sequence shown in Fig. \ref{MQCp}) along with a reference spectrum are shown in Fig. \ref{TMP}.  
  
\begin{figure}[t]
 	\includegraphics[trim = 1cm 1cm 1cm 1cm, clip, width=8.5cm ]{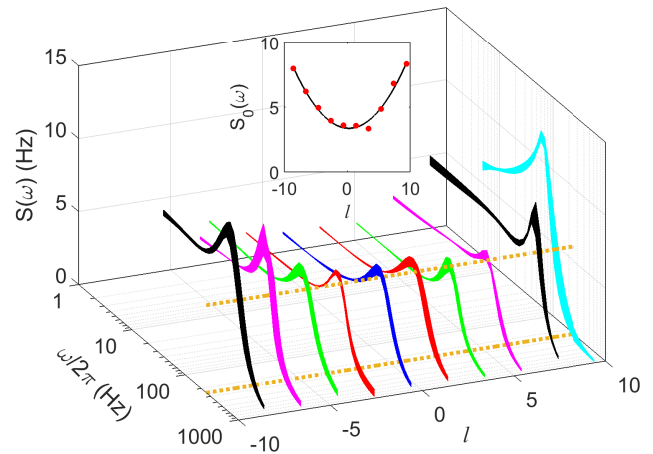} 
 	\caption{Trimethylphosphite noise-spectra for various MSSM states with different lopsidedness $l$. The dashed lines parallel to $l$-axis represent the maximum frequency (250 Hz) sampled in experiments. The inset shows the scaling of low-frequency spectral density values with $l$.}
 	\label{MQEns}
 \end{figure}

\textit{Results and discussions:}
The results of the noise spectroscopy of various MSSM states are shown in Fig. \ref{MQEns}.  As expected, the spectral density profiles appear to go higher with the magnitude of the lopsidedness, and accordingly the NOON state has the highest noise profile.  

It is interesting to study the scaling of the low-frequency noise ($\approx S(0)$) versus the lopsidedness.
The inset of Fig. \ref{MQEns} shows the experimental values of low-frequency noise (at lowest frequencies sampled) and a fit with a shifted parabola $c_2 l^2 + c_0$.  The best fit was found at $c_2 = 0.06\pm0.01 $ and $c_0 = 3.37 \pm 0.34$. 
A quadratic scaling of noise with lopsidedness is obvious from the inset in Fig. \ref{MQEns}.

According to Redfield theory of relaxation, the transverse relaxation is a result of two processes - adiabatic and nonadiabatic \cite{redfield, Abragam}.  The energy conserving adiabatic part arises by longitudinal noise and leads to dephasing. The nonadiabatic part is due to the transverse noise and can induce transitions.  Tang et al had observed that the completely correlated longitudinal noise results in relaxation rates that vary quadratically with the coherence order \cite{pines}.  In our system, the coherence order is characterized by  lopsidedness.  Thus the quadratic dependence of spectral density with lopsidedness points out that the noise is predominantly correlated, i.e., noise affects all the spins identically.  The background part in the scaling ($c_0$) is due remaining contributions including the nonadiabatic relaxation and the self-relaxation of the probe qubit ($^{31}$P).

It can be noted that similar studies of scaling of decoherence were earlier reported in a solid state NMR system by Krojanski et al  \cite{krojanski2004scaling}.
  
  \begin{figure}[b]
  	\includegraphics[trim = 0.5cm 9.5cm 0cm 9.5cm, clip, width=8cm ]{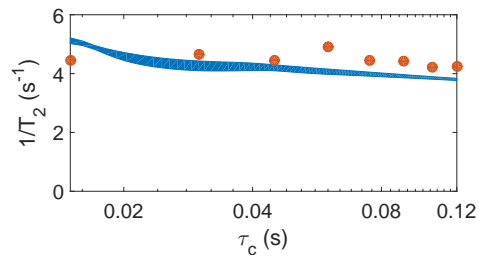} 
  	\caption{Decay rates versus UDD-3 cycle duration $\tau_c$ calculated using Eq. \ref{eq7} for the experimental noise spectrum of 10-qubit NOON state.  The dots correspond to experimental results.} 
  	\label{DDcomp}
  \end{figure}

An immediate application of extracting the noise spectrum is in evaluating the performances of various types DD sequences and selecting the optimum sequence for  preserving quantum coherences.  Uhrig dynamical decoupling (UDD) \cite{Uhrig}, for example involves, a nonuniform distribution of  $\pi$ pulses placed at time instants
\begin{eqnarray}
t_j = \tau_c \sin^2 \left( \frac{\pi j}{2N_\pi +2} \right),
\end{eqnarray}
where $N_\pi$ is the total number of $\pi$ pulses in one period ($\tau_c$), also known as the order of the UDD sequence (denoted UDD-N$_\pi$).
It can be easily seen that UDD-1 and UDD-2 are identical to a CPMG sequence.  

Having the functional form of the
noise spectral density we can now predict the relative decay rates of a quantum state under a given DD sequence.  As an example, the band in Fig. \ref{DDcomp} shows the the predicated decay rates of the NOON state (spectral density shown in Fig. \ref{MQEns}) under UDD-3 sequence for a range of $\tau_c$ values.  The corresponding experimental decay rates are shown by dots.  The reasonable agreement between experimental and predicted values of decay rates demonstrates the benefit of extracting the spectral distribution of noise.  Similar results were obtained in the case of other MSSM states.  It should be noted that imperfections in the $\pi$-pulses such as finite duration, sensitivity to RF inhomogeneity over the sample volume, and calibration errors may introduce additional uncertainties in the noise-spectrum estimation and may affect DD performance as well.

%%%%%%%%%%%%%%%%%%%%%%%%%%%%% Conclusion %%%%%%%%%%%%%%%%%%%%%%%%%%%%%%%%%%%%%%%%%%%%%%%%

\section{Conclusions}
While we are entering the era of quantum devices, noise remains a hurdle in storing quantum superpositions.  Exploiting decoherence-free-subspaces (DFS) is one of the convenient ways to preserve quantum coherences. 
DFS is already being used for storing hyper-polarization \cite{ahuja2010proton}, studying slow molecular dynamics \cite{sarkar2007singlet}, characterizing molecular diffusion \cite{sarkar2008measurement,cavadini2005slow}, precise measurements of coupling constants \cite{bornet2011ultra}, as well as in fault-tolerant quantum computing \cite{bacon2000universal}.  
However the noises influencing such special quantum coherences have not been hitherto characterized experimentally.  In this work we have experimentally characterized  and compared noise spectral densities of various multi-qubit coherences.

We found that the noise spectrum of the long-lived singlet state (LLS) under spin-lock of a homonuclear spin-pair had the lowest profile indicating the strong protection offered by the symmetry in DFS resulting in long-livedness of the state.  The long-lived-coherence (LLC) between singlet and the zero-quantum triplet had a higher noise profile, but still lower than the normal uncorrelated (single-spin) coherence. 
We have also measured the extent of noise in LLS under different field strengths and as expected, we found a higher noise with a stronger field, although the overall spectral features remained similar.  On the other hand, the uncorrelated spins showed lower noise content compared to singlet states in a heteronuclear spin system, indicating an asymmetry in the system.  Further, we have also explored the noise profiles of various higher-order coherences in a 10-spin system, and found a predominantly quadratic scaling of noise with respect to coherence order. Finally, using the noise spectrum of the NOON state we predicted its decay rates under a 3rd order Uhrig dynamical decoupling sequence and verified the same with experiments. 

We believe that such studies are useful for understanding the physics of noise affecting quantum systems as well as to design ways to suppress decoherence.  A better understanding of noise and their suppression will be crucial not only for the physical realization of quantum devices but also for general spectroscopic applications.

\section*{Acknowledgements}
We acknowledge useful discussions with Swathi Hegde and Abhishek Shukla.  This work was supported by DST/SJF/PSA-03/2012-13 and CSIR 03(1345)/16/EMR-II.
%Authors dedicate this work to Prof. Anil Kumar on his 75th year.

\section*{References}
  \bibliography{NScite}{}
  \bibliographystyle{unsrt}

  \end{document}